\documentclass [11pt,notoc]{article}
\usepackage{jheppub}

\newcommand{\bea}{\begin{equation} \begin{aligned}} 
\newcommand{\eea}{\end{aligned} \end{equation}}

\def\beq{\begin{equation}} \def\eeq{\end{equation}}

\usepackage[english]{babel}
\usepackage{csquotes}
\MakeOuterQuote{"}

\notoc

\makeatletter
\def\@fpheader{\ }
\makeatother

\title{Comment on "Redundancy Channels in the Conformal Bootstrap" by S. R. Kousvos and A. Stergiou}
\author{Slava Rychkov}

\affiliation{Institut des Hautes \'Etudes Scientifiques, 91440 Bures-sur-Yvette, France}
\emailAdd{slava@ihes.fr}
	
\abstract{Recent work by Kousvos and Stergiou criticises our work with Zhong Ming Tan \cite{Rychkov:2015naa}. The issue is CFT scaling dimension computations in perturbative Renormalization Group. We identified operators whose correlation functions differ by contact terms. This is allowed because CFT only describes correlation functions away from coincident points. They instead renormalize the contact terms, which are eventually dropped. Our way is not only correct, but preferable as it operates with the minimal set of quantities. }

\begin{document}
	\maketitle

\newpage

Consider the Wilson-Fisher fixed point, i.e.~the IR fixed point of the $\phi^4$ theory in $d=4-\epsilon$ dimensions.  This IR fixed point is a CFT, and scaling dimensions of CFT operators can be computed in perturbation theory. In doing so, one has to face various well-known technical issues, such as e.g.:
\begin{enumerate}
	
\item One needs to renormalize, in a chosen scheme. Renormalization makes it manifest that the fixed point is weakly coupled, because the renormalized coupling is order $\epsilon$, while the bare coupling may be order one. Most authors use the minimal subtaction (MS) scheme where the theory is renormalized by subtracting poles in $\epsilon$; see e.g.~\cite{Vasil'ev1998,Kleinert2001}. 

\item Composite operators also need to be renormalized. The bare operators $\mathcal{O}_i$ are related to the renormalized operators $[\mathcal{O}_i]$ by a mixing matrix $Z_{ij}(\lambda;\epsilon)$ which contains a series of ascending poles in $\epsilon$, with coefficients depending on the coupling. Eigenvalues of the derivative of the mixing matrix determine the anomalous dimensions of the operators at the IR fixed point, while the eigenvectors give the corresponding operators. In the MS scheme, the mixing matrix has a block-diagonal form as only operators having the same classical scaling dimension in the $\epsilon\to0$ limit may mix.

\end{enumerate}
There is a third issue, which motivates this comment: the role of equations of motion (EOM). Correlation functions of any operator $\mathcal{O}_{\rm EOM}(x)$ proportional to the EOM are contact terms, i.e.~proportional to (derivatives of) delta-functions $\delta^{(d)}(x-x_i)$, where $x_i$ are coordinates of other operators. As such, these EOM operators do not belong to the CFT, which only describes the operators whose correlation functions are nonzero at non-coincident points. 

In regard to this issue, there are two ways to proceed:
\begin{enumerate}
\item (Full way) One renormalizes the full list of composite operators of a given classical dimensions, including those proportional to the EOM. In the end, EOM operators are dropped, and the remaining ones retained, since only the latter scaling dimensions are a part of the CFT data. This is the way that Kousvos and Stergiou (K\&S) \cite{Kousvos:2025ext} used in their work.

\item (Economical way) One identifies operators differing by something proportional to the EOM. In this approach, one renormalizes correlation functions of composite operators modulo contact terms. Since CFT only cares by non-coincident points, one is guaranteed to get correct results for CFT dimensions. This is the way one normally uses in a practical computation, since it avoids unnecessary work.
\end{enumerate}
The simplest example is the mixing of $\phi^3$ and $\partial^2\phi$ operators. In the free theory, $\partial^2\phi=0$ modulo contact terms, while $\phi^3$ is a non-trivial operator. At the Wilson-Fisher fixed point, both $\phi^3$ and $\partial^2\phi$ are nontrivial, but they are related by the EOM, so away from coincident points there is again just one one independent operator. In our work \cite{Rychkov:2015naa}, $\phi^3$ and $\partial^2\phi$ were treated as identical (up to rescaling) operators at the Wilson-Fisher fixed point, in particular having equal scaling dimensions. We stand by this statement, understanding the scaling dimension in the CFT sense, i.e. from the scaling of correlation functions away from the coincident points. K\&S, instead, use the term "scaling dimension" in the strict perturbative RG sense, considering the full mixing matrix, including contact terms. In their language $\phi^3$ would not even be considered a scaling operator, since under scale transformations its correlation functions transform nonhomogeneously. However, there is no contradiction with our work since the nonhomogeneous contribution is a contact term, invisible in the CFT.

A similar discussion can be made for other RG flows close to the upper or lower critical dimensions, such as multiscalar fixed points in $d=4-\epsilon$. E.g.~K\&S discussed a conserved current $J_\mu$ of the $O(N)$ fixed point which eats a quartic operator $\mathcal{O}_4$ and becomes non-conserved. K\&S consider the difference $\partial_\mu J_\mu - \mathcal{O}_4$ as a scaling operator proportional to EOM. We would instead say that away from coincident points $\partial_\mu J_\mu$ and $\mathcal{O}_4$ are both scaling operators, which are identified thanks to EOM. 

Unfortunately, K\&S's comments about our work make it seem that the difference of treatment is substantial. On the contrary, we believe it is purely terminological, without consequences for CFT scaling dimensions. Moreover, our way of thinking, speaking and computing is more economical. We are not aware of any reason not to use it in the original context of \cite{Rychkov:2015naa}. It could have also perfectly fit the needs of \cite{Kousvos:2025ext}. 


\providecommand{\href}[2]{#2}\begingroup\raggedright\endgroup

\end{document}